\documentclass[aps,prl,reprint,superscriptaddress,nobibnotes,nofootinbib]{revtex4-2}
\usepackage[utf8]{inputenc}

\usepackage{xcolor}
\usepackage{amsmath,amssymb,bm}
\usepackage{amsfonts}
\usepackage{graphicx}
\usepackage{tikz}
\usepackage{siunitx}
\usepackage{mathtools}

\DeclarePairedDelimiter{\ket}{\lvert}{\rangle}
\usepackage[colorlinks=true,allcolors=blue]{hyperref}
\usepackage[capitalise,noabbrev]{cleveref}

\begin{document}
\title{Fundamental Limits of Large Momentum Transfer in Optical Lattices}

\author{Ashkan Alibabaei}
\thanks{These authors contributed equally to this work.}
\affiliation{Institute of Quantum Optics, Leibniz University of Hannover, Welfengarten 1
30167 Hannover, Germany}

\author{Patrik Mönkeberg}
\thanks{These authors contributed equally to this work.}
\affiliation{Institute of Quantum Optics, Leibniz University of Hannover, Welfengarten 1
30167 Hannover, Germany}
\affiliation{Institute of Theoretical Physics, Leibniz University of Hannover, 
Appelstrasse 2
30167 Hannover, Germany}
\affiliation{Institute for Theoretical Physics, University of Innsbruck, 6020 Innsbruck, Austria}

\author{Florian Fitzek}
\thanks{Currently at ASML Netherlands B.V., De Run 6501, 5504 DR Veldhoven.}
\affiliation{Institute of Quantum Optics, Leibniz University of Hannover, Welfengarten 1
30167 Hannover, Germany}
\affiliation{Institute of Theoretical Physics, Leibniz University of Hannover, 
Appelstrasse 2
30167 Hannover, Germany}

\author{Michael Werner}
\affiliation{Institute of Quantum Optics, Leibniz University of Hannover, Welfengarten 1
30167 Hannover, Germany}

\author{Alexandre Gauguet}
\affiliation{Laboratoire Collisions Agrégats Réactivité, FERMI, Université de Toulouse and CNRS UMR5589, 118 Route de Narbonne, F-31062 Toulouse, France}

\author{Baptiste Allard}
\affiliation{Laboratoire Collisions Agrégats Réactivité, FERMI, Université de Toulouse and CNRS UMR5589, 118 Route de Narbonne, F-31062 Toulouse, France}

\author{Klemens Hammerer}
\email[Contact author: ]{klemens.hammerer@uibk.ac.at}
\affiliation{Institute of Theoretical Physics, Leibniz University of Hannover, 
Appelstrasse 2
30167 Hannover, Germany}
\affiliation{Institute for Theoretical Physics, University of Innsbruck, 6020 Innsbruck, Austria}
\affiliation{Institute for Quantum Optics and Quantum Information of the Austrian Academy of Sciences, 6020 Innsbruck, Austria}

\author{Naceur Gaaloul}
\email[Contact author: ]{gaaloul@iqo.uni-hannover.de}
\affiliation{Institute of Quantum Optics, Leibniz University of Hannover, Welfengarten 1
30167 Hannover, Germany}

\date{\today}
\begin{abstract}
Large-momentum-transfer techniques are instrumental for the next generation of atom interferometers as they significantly improve their sensitivity. State-of-the-art implementations rely on elastic scattering processes from optical lattices such as Bloch oscillations or sequential Bragg diffraction, but their performance is constrained by imperfect pulse efficiencies. Here we develop a Floquet-based theoretical framework that provides a unified description of elastic light–atom scattering across all relevant regimes. 
Within this formalism, we identify practical regimes that exhibit orders of magnitude reduced losses and improved phase accuracy compared to previous implementations. The model's validity is established through direct comparison with numerical solutions of the Schrödinger equation and by quantitative agreement with recent experimental benchmark results. These findings delineate previously unexplored operating regimes for large momentum transfer beam splitters and open new perspectives for precision atom-interferometric measurements in fundamental physics, gravity gradiometry or gravitational wave detection.
\end{abstract}
\maketitle
\begin{figure}
    \centering
    \includegraphics[width=1.0\linewidth]{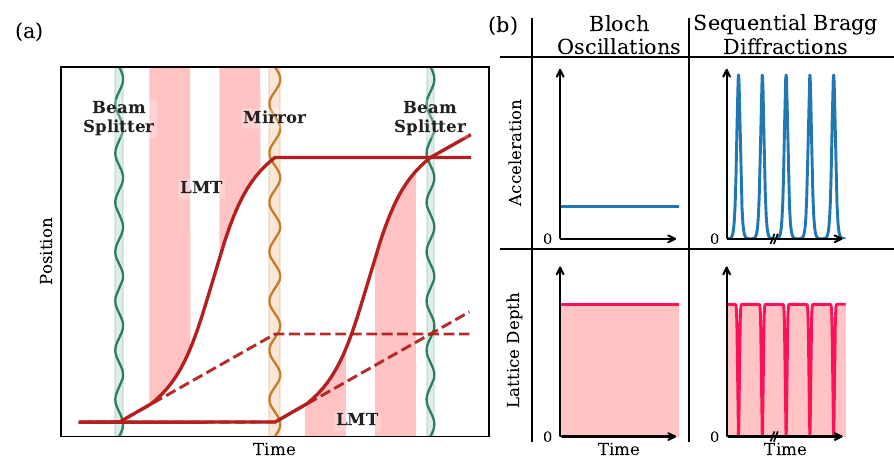}
    \caption{(a) Space-time diagram of a Large-Momentum-Transfer-enhanced Mach-Zehnder atom interferometer in the freely falling frame. The sequence consists of two beam splitter and a mirror pulse. Each interferometer arm is accelerated and decelerated using LMT sequences in an optical lattice, resulting in a larger space-time area covered compared to a standard Mach-Zehnder atom interferometer (dashed lines). (b) Control parameters (acceleration and lattice depth) for Bloch Oscillation and Sequential Bragg Diffraction LMT.}
    \label{fig:1}
\end{figure}

\textit{Introduction---}
Atom Interferometry has proven to be a highly sensitive and accurate tool for quantum sensing~\cite{initial_2} advancing inertial navigation~\cite{initial_effect,Kalman_filter}, precision measurements~\cite{raman_kasevich,g_measurement_kasevich,g_measure,Geophy}, fundamental constants determination~\cite{Lamporesi_G_constant,Parker_muller,Nature_fine_Structure_constant} and tests of general relativity~\cite{GR1,GR2}. 
One way to enhance the sensitivity of an atom interferometer is to increase the spatial separation between the arms with sizable photon exchange, a technique known as Large Momentum Transfer (LMT)~\cite{LMT1,LMT2,LMT3,LMT4}. Methods to realize LMT include single photon transitions~\cite{one_photon_first,LMT_singlephoton,one_photon_rudolph}, Raman diffraction~\cite{raman_3,Raman2}, as well as elastic-scattering in optical lattices such as Bloch oscillation (BO)~\cite{Bloch_oscillation_peik,Bloch1,Bloch2,Gebbe} and Sequential Bragg Diffraction (SBD)~\cite{Toulouse1,Toulouse2}. 
Recently, momentum separations of up to $600$ photon recoils have been achieved using SBD~\cite{Toulouse2}, exceeding earlier $400$ photon recoils in BO-based implementations~\cite{Gebbe}. These results raise the following questions: (i) whether BO and SBD constitute fundamentally distinct elastic scattering mechanisms as illustrated in Fig.~\ref{fig:1}; (ii) how their efficiency and accuracy compare under experimental constraints; and (iii) whether either approach can achieve momentum separations of several thousand photon recoils as required for next-generation atom-interferometric sensors detecting gravitational waves or investigating the nature of dark energy and dark matter~\cite{GW_detection_ai_massimo,gw_kasevich,MIGA,ELGAR,ZAIGA,MAGIS,AION,TVLBAI2}.
In this work, we exploit the temporal periodicity and discrete spatial symmetry of optical lattices within a Floquet framework to extend a recently developed Wannier–Stark model~\cite{Bloch_paper} to the full class of elastic LMT scattering methods. Within this unified description, we introduce a parametrized family of Hamiltonians that continuously interpolates between the limiting cases of BOs and SBDs through variation of a single control parameter. This approach exposes the fundamental efficiency and phase-accuracy limits inherent to different LMT schemes and enables their direct comparison on equal footing. Furthermore, it identifies previously unexplored operating regimes in which pulse efficiencies can be substantially enhanced, thereby extending the attainable performance of state-of-the-art implementations by several orders of magnitude. 
The paper is structured as follows: We first present the theoretical framework by introducing the general Hamiltonian and presenting a solution of the time-dependent Schrödinger equation based on Floquet theory. We then proceed to parametrize a specific family of Hamiltonians to cover BOs and SBDs as two distinct limits which allows us to evaluate our results by analyzing the efficiency and phase accuracy, drawing comparisons with state-of-the-art experiments. Then, we explore the consequences of our approach through its application to a toy model of an LMT-enhanced folded gradiometer, highlighting the emergence of a fundamental sensitivity limit.  Finally, we present a method for adiabatic preparation of Floquet states in optical lattices and compare our results to exact numerical solutions of the Schrödinger equation.

\textit{Theoretical model---}
We consider an atom with a mass $m$ loaded in an optical lattice which is formed by two counter-propagating beams characterized in the laboratory reference frame by their phases \(\phi_{1,2}\), their frequencies \(\omega_{1,2}\), and their opposite wave vectors \(\mathbf{k}_{1,2}\) with \(\mathbf{k}_1 \approx -\mathbf{k}_2\). The phase and frequency differences between these two beams are defined as \(\phi(t) = \phi_1(t) - \phi_2(t)\) and \(\omega_L(t) = \partial_t\phi \), respectively. The mean wave vector is given by \(k_L = (k_1 - k_2)/2=\pi/d\), where $d$ is the lattice period. When the two beams overlap, they give rise to a quasi-stationary wave moving with acceleration \(a_L(t) = \dot\omega_L(t) /2k_L\) relative to the laboratory reference frame. The laser is detuned far from the frequencies of the internal atomic transitions, allowing for adiabatic elimination of the excited state. This leads to an interaction potential of the form $V(t)\cos^2\left(k_L\hat{x} - \ \phi(t)/2\right)$, with lattice depth $V(t)$. In the following we consider time-periodic lattice parameters $V(t+ T_F) = V(t)$ and $a_L(t+T_F) = a_L(t)$, where $T_F$ is the Floquet period. This includes SBDs, where the lattice depth is constant and the lattice acceleration $a_L(t)=a_0\sum_{n}\delta(t-nT_F)$ is a Dirac-comb, as well as BOs, where both lattice depth and acceleration $a_L(t)=a_0$ are constant. We relate the Bloch acceleration $a_0=2\hbar k_L / m T_B$ to the Floquet period by setting $T_B = T_F$, where $T_B$ is the Bloch period; see Supplemental Material at section I for more details~\cite{supp_mat}.
In the co-moving \emph{lattice frame}~\cite{Bloch_oscillation_peik,supp_mat}, the Hamiltonian takes the form
\begin{align}\label{eq:Hlatframe}
    H(t)=\frac{\hat{p}^2}{2m}+V(t)\cos^2(k_L\hat{x})+ma_L(t)\hat{x}
\end{align}
with time-periodicity \(H(t+T_F)=H(t)\). A solution to the time-dependent Schrödinger equation can be obtained by diagonalizing the Floquet operator $U(T_F)$, that is, the time-evolution operator over one period~\cite{Holthaus_2016}. Here, the Floquet operator has a discrete spatial symmetry under translations by $d$; see section I of Supplemental Material~\cite{supp_mat}. This implies that $U(T_F)$ and the discrete spatial translation operator $\hat{T}$ share a common eigenbasis, the Floquet states, defined by 
\begin{align}\label{eq:eigenequation_floquetop}
    U(T_F)\,\ket{\phi_\alpha(\kappa)}&=e^{-i \mathcal{E}_\alpha(\kappa)T_F/\hbar}\ket{\phi_\alpha(\kappa)}, \\
    \hat{T}\,\ket{\phi_\alpha(\kappa)}&=e^{i\kappa d}\ket{\phi_\alpha(\kappa)}
\end{align}
with quasi-momentum $-k_L\leq\kappa\leq k_L$ and a band index $\alpha$. The quasi-energies in general depend on the quasi-momentum $\kappa$ and are complex numbers of the form
$\mathcal{E}_{\alpha, \ell}(\kappa) = E_\alpha(\kappa) + 2\pi\ell \hbar / T_F -i\Gamma_\alpha(\kappa)/2
$~\cite{Complex_energy}.
The imaginary part of the energies represents the linewidth of the Floquet state, determining the tunneling losses to the continuum~\cite{Bloch_paper,Bloch_Wannier,Complex_energy} in the tilted potential of Eq.~\eqref{eq:Hlatframe}. These losses result from the finite spatial extent of the system, in contrast to ideal Floquet states, which are infinitely extended and thus not physically realizable.
Due to the periodicity of the complex exponential, each quasi-energy corresponds to a set of equivalent representatives given by $\ell \in \mathbb{Z}$, where the quantum number $\ell$ can be associated with the lattice site. See section II of Supplemental Material~\cite{supp_mat} for more details on the dependence of the quasi-energies $\mathcal{E}_\alpha(\kappa)$ on the quasi-momentum. In the following, we analyze the efficiency of LMT schemes in terms of their quasi-energies and, in particular, their intrinsic losses quantified by $\Gamma_\alpha(\kappa)$.

\textit{Unifying Bloch and sequential Bragg LMT---} We consider a family of pulses that allows for a smooth interpolation between BOs and SBDs. Subsequently we assume a constant lattice depth $V(t)=V_0$ and parametrize the lattice acceleration accordingly with an interpolation parameter $\eta\in(0,\infty)$. On the level of the lattice momentum, to interpolate between a linear profile (BOs) and a Heaviside-step function (SBDs), we employ a smooth sigmoid function. This corresponds to the lattice acceleration:
\begin{align}\label{eq:lattice_param}
   a_L^{(\eta)}(t) &= \frac{1 + \cosh(\eta/2)}{\sinh(\eta/2)} \frac{a_0\eta e^{-\eta(t/T_F - (n + 1/2))}}{\big(e^{-\eta(t/T_F - (n + 1/2))}  + 1\big)^2},
\end{align}
where $nT_F \leq t < (n+1)T_F$, $a_0 = 2\hbar k_L / m T_F$. This is a mere example of a possible parametrization family for interpolating between BOs and SBDs. The quasi-momentum $\kappa$ remains as a degree of freedom of the system, relevant for the LMT efficiency and is fixed via the initial Floquet state. For separate Bragg pulses the quasi-momentum is related to the Bragg resonance condition which is fulfilled at $\kappa = k_L$. However for the case of sequential pulses the quasi-momentum can be chosen freely without taking the resonance condition into account. Moreover, we later show that the choice of $\kappa= 0$ can result in a much higher efficiency of the corresponding LMT sequence. Even though the Bragg resonance condition is in general not fulfilled for a freely chosen quasi-momentum, we will here call the entire regime of sequential delta-like acceleration pulses ``SBD''.
\begin{figure}
    \centering
    \includegraphics[width=1.0\linewidth]{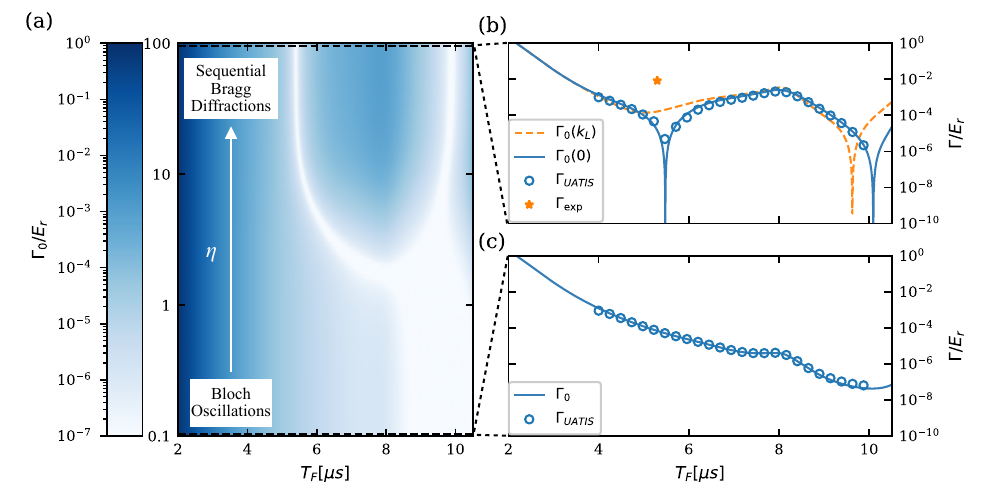}
    \caption{(a) Density map of the tunneling loss rate $\Gamma_0$ of the Floquet-groundstate at zero quasi-momentum, $\vert \phi_0(0)\rangle$, versus Floquet period $T_F$ and interpolation parameter $\eta$ (cf. Eq.~\eqref{eq:lattice_param}) for a lattice depth of $V_0 = 50E_r$, where $E_r=\hbar^2 k_L^2/2m$ is the recoil energy. The color bar is cut off at $10^{-7}$ to increase the contrast. (b) Cut through the density map (a) in SBD limit at $\eta=100$ (solid blue line) corresponding to $\kappa=0$, results of exact numerical simulations (blue points), tunneling loss rate in SBDs limit for $\kappa=k_L$ (orange line) and overall loss rate of state-of-the-art experiment~\cite{Toulouse2} (orange star). The experimental loss rate deviates from its fundamental limit at $\kappa=k_L$ due pulse-to-pulse fluctuations and spontaneous emission losses. (c) Cut through the density map (a) in BO limit at $\eta=0.1$ (solid blue line) and results of exact numerical simulations (blue points). The exact numerical simulations are performed with the ``Universal Atom Interferometer Simulator'' (UATIS)~\cite{UATIS}.}
    \label{fig:2}
\end{figure}
Fig.~\ref{fig:2}\,(a) illustrates the smooth interpolation between BOs and SBDs characterized by the linewidth $\Gamma_0$ of the Floquet state $\ket{\phi_0(\kappa=0)}$ of the fundamental band $\alpha=0$. In general, the linewidth $\Gamma_\alpha$ corresponds to the tunneling loss rate of the $\alpha$-th Floquet state to the continuum, either directly or via intermediate Floquet states. In the BO regime, the spectrum resembles those previously reported in~\cite{Bloch_paper}, where the peaks in the tunneling loss rate can be explained via tunneling resonances between different Wannier-Stark ladders~\cite{Bloch_paper}. Remarkably, beyond a certain threshold for $\eta$, one observes sharp \emph{anti-resonances} in the line width $\Gamma_0$, corresponding to a pronounced suppression of tunneling losses. This phenomenon is reminiscent of coherent suppression of tunneling~\cite{CDT} and dynamical localization in shaken optical lattices~\cite{dynamical_localisation,dynamical_localisaion_2,shaken_1,shaken_lattice_2,shaken_lattice_3}, which also occurs for excited bands; see section III of Supplemental Material~\cite{supp_mat}. Figs.~\ref{fig:2}\,(b,c) show two representative cuts through the density map Fig.~\ref{fig:2}\,(a), obtained by fixing the parameter $\eta$ to values corresponding to the BO and SBD regimes, respectively. At anti-resonances the SBD regime shows a clear enhancement in fundamental efficiency compared to the BO case for the same periods $T_F$, thus allowing for more efficient LMT at fast transfer rates. 
Fig.~\ref{fig:2}\,(b) additionally highlights the crucial role of the quasi-momentum, revealing a clear improvement from states prepared at $\kappa=k_L$ to states prepared at $\kappa=0$ at short Floquet periods. While $\kappa=k_L$ might be the more intuitive choice as mentioned before, its first anti-resonance occurs at a significantly longer Floquet period compared to $\kappa=0$. Consequently, achieving comparable efficiency would correspond to almost half the momentum transfer rate. Next to tunneling losses, spontaneous emission losses must be accounted for, where a sufficiently high laser power is required to resolve the fundamental efficiency associated with the anti-resonances; see section IV of Supplemental Material~\cite{supp_mat}.
Finally, we analyze one of the main contributors to the overall phase uncertainty, namely that induced by lattice intensity asymmetries between the interferometer arms and pulse-to-pulse intensity fluctuations within a single interferometer arm~\cite{Toulouse2}. We find no qualitative difference in this dephasing between the BO and SBD regimes, indicating that the energies $E_{\alpha,\ell}$ depend only on the average acceleration and not on its specific modulation; see section V of Supplemental Material~\cite{supp_mat} for more details on the phase uncertainty. Notably, at certain lattice depths the phase uncertainty vanishes for specific excited states. In the BO case this phenomena was previously reported in~\cite{Gupta} as  ``magic'' lattice depths, defining dephasing-robust subspaces. We report the existence of these subsystems robust to dephasing in the general case, opening the possibility for optimization, where high-fidelity pulses can also be robust to dephasing.

\begin{figure}
    \centering
    \includegraphics[width=1.0\linewidth]{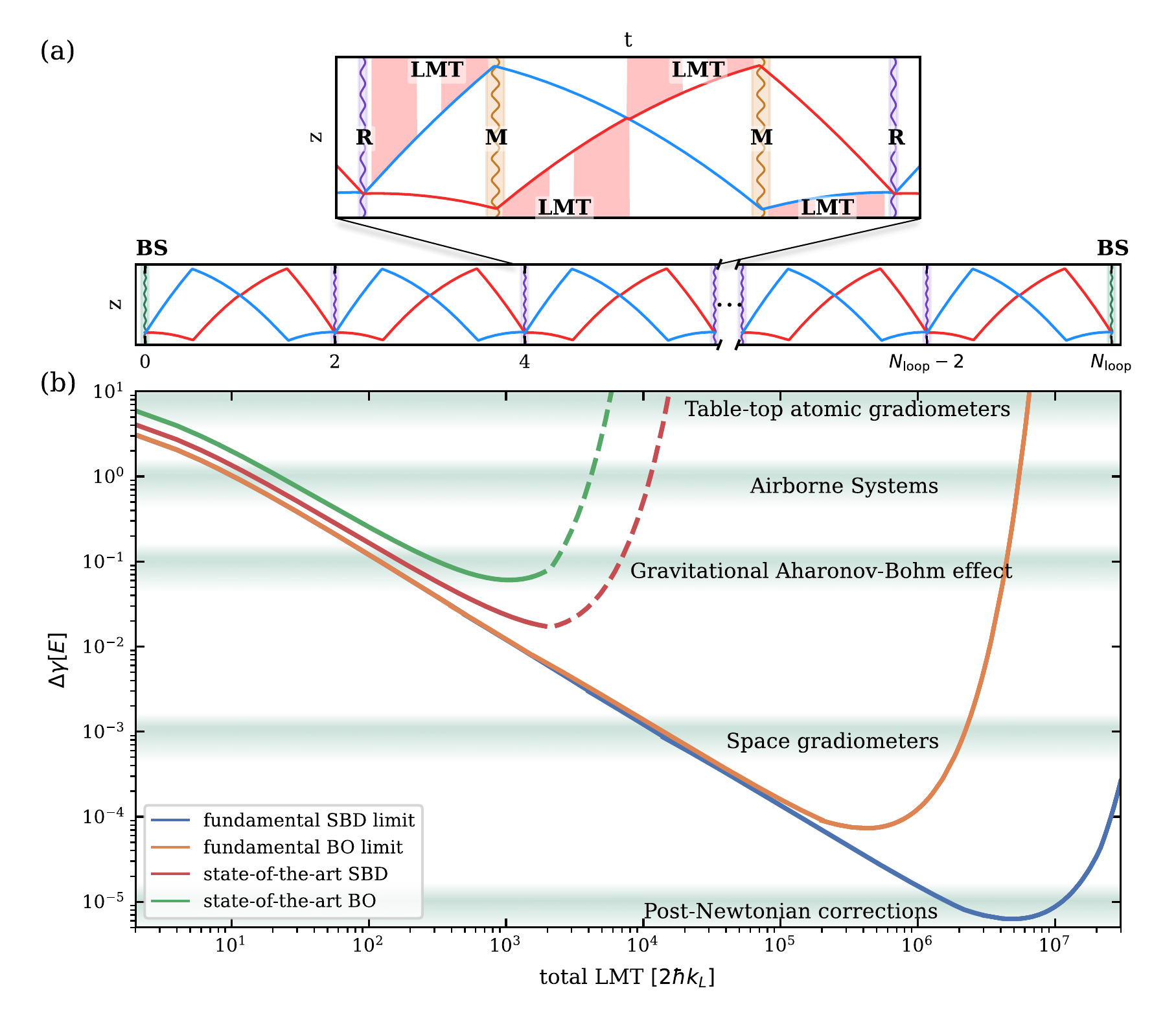}
    \caption{(a) Schematic of LMT-enhanced multi-loop interferometer sequence, including beam splitter pulses before the first loop and after the last loop. Additional relaunching pulses are applied at the end of every second loop. The inset shows a detailed view of two consecutive loops, including two relaunch pulses (R), two mirror pulses (M) and four LMT sequences (LMT).
    (b) Gravity gradient sensitivity $\Delta \gamma$ (in Eötvös, $1E=10^{-9}\, \mathrm{s^{-2}}$) as a function of LMT order $N_\text{LMT}$ for a $5$-$\mathrm{m}$-tall baseline. The curves show the shot-noise-limited sensitivity for state-of-the-art BO~\cite{Gebbe} and SBD~\cite{Toulouse2} efficiencies extrapolated to the presented geometry, as well as their corresponding fundamental limits at a Floquet period of $T_F= 5.5\,\mu\mathrm{s}$ with a lattice depth of $V_0=50E_r$ including spontaneous emission for a laser system with a power of $20\,\mathrm{W}$ and a beam waist of $w=1.6\,\mathrm{mm}$. An initial atom number of $N_\text{atoms} = 10^9$ of $^{87}\mathrm{Rb}$, with a momentum width of $\Delta_p \leq 0.2\hbar k_L$ and a free evolution time of $T_\text{free}=0.3\,\si{s}$ between acceleration and deceleration, is assumed. The drift time and folded geometry restrict the baseline to a  maximum of few meters. The dashed parts of the curves point out arm separations exceeding the set limit of $5\,\si{m}$. Shaded bands indicate representative sensitivity benchmarks for table-top quantum gradiometers~\cite{Quantum_gradiometers}, airborne geophysical systems~\cite{Airborne}, the resolution required to observe the gravitational Aharonov-Bohm effect~\cite{ABeffect}, the ESA gradiometry mission GOCE~\cite{ESA_GOCE}, and post-Newtonian corrections to gravity~\cite{MW_curvature,GR_Test,Alibabaei_2023}.}
    \label{fig:3}
\end{figure}

\textit{Application---}
Among the various applications of LMT-enhanced atom interferometry, we choose here gradiometry as a study case. Specifically, we consider an LMT-enhanced folded geometry characterized by the pulse sequence $\frac{\pi}{2}-\pi-\dots-\pi-\frac{\pi}{2}$, with even number of $\pi$ pulses, as illustrated in Fig.~\ref{fig:3}\,(a). Multi-loop geometries have been proposed~\cite{multiloop_relaunch_Schubert,Schach_2025} and experimentally realized~\cite{double_loop_gradiometry,multipath_experiment,200_loops} in several contexts, including implementations with large numbers of pulses and hundreds of interferometric loops. While the specific scheme is not central to our conclusions, it illustrates how the optimized Floquet-state basis identified here could enable momentum transfer at the level of several million photon recoils in practical configurations suitable for medium-to-long-baseline interferometers. 
To keep the interferometer baseline at a constant height, we consider a relaunch of the atoms after every second loop, where both interferometer arms are vertically relaunched simultaneously as previously proposed in~\cite{multiloop_relaunch_Schubert}. Experimentally, a similar relaunch has been realized with one optical lattice in~\cite{Atomchip_relaunch_Abend} or in a twin-lattice approach in reference~\cite{Gebbe}. We consider the same pulse types in the relaunch as the respective LMT sequences, allowing a fair efficiency comparison across regimes. To avoid cross-talk between the interferometer arms during the relaunch, a minimal momentum separation between the interferometer arms has to be created using high-order Bragg double-diffraction as implemented in reference~\cite{Gebbe}.
Assuming operation at the shot-noise limit, the corresponding gravity gradient sensitivity is given by
\begin{align}\label{eq:sens}
    \Delta \gamma \ge \frac{1}{\mathcal{Q}}\frac{1}{\sqrt{N_\text{atoms}}}\frac{1}{N_\text{loops}},
\end{align}
where $\gamma$ denotes the gravity gradient, $N_\text{atoms}$ the number of detected atoms~\cite{Shotnoise_Gauguet,shot_noise,shot_noise_2}, and $\mathcal{Q}$ is the scale factor specific to the interferometer geometry, scaling with the momentum transfer per loop $N_\text{LMT}^\text{loop}$. A more detailed derivation of the sensitivity, comparisons to numerical simulations as well as relaunch scheme can be found in section VI of the Supplemental Material~\cite{supp_mat}.
As discussed in previous sections, the number of detected atoms is fundamentally constrained by the fundamental losses to the continuum and spontaneous emission losses. Fig.~\ref{fig:3}\,(b) illustrates the scaling of the gravity gradient sensitivity with the amount of LMT ($2\,\hbar k_L$ per pulse) for a compact setup (limited to few meter-high fountains) with a given height set by the maximum arm separation. We increase the total LMT order $N_\text{LMT}^\text{total} = N_\text{loop}\cdot N_\text{LMT}^\text{loop}$ by first increasing the momentum transfer per loop up to a maximum of $N^\text{loop}_\text{LMT}=1000$ and subsequently increasing the number of loops $N_\text{loop}$, which initially enhances the sensitivity according to Eq.~\eqref{eq:sens}. Beyond a certain total LMT order, set by the efficiency of the chosen LMT method, the decreasing atom number due to fundamental losses and spontaneous emissions leads to increased shot-noise that dominates the sensitivity. This leads to an maximum total LMT order, up to which the sensitivity improves before diverging. Higher efficiencies 
of the LMT sequence, as identified in this work, dramatically extends the number of useful photon recoils, as reflected by the respective minima in Fig.~\ref{fig:3}\,(b) reaching a million for the BO case and $10^7$ in the SBD. This performance will allow to explore a wealth of physics cases across multiple disciplines such as fundamental physics, Earth sciences or gravitational wave detection. Note that the folded geometry presented in Fig.~\ref{fig:3}\,(a) is not optimized for a given measurement but merely serves as a proof-of-concept for showcasing the possibilities opened by applying the optimized LMT sequences at the fundamental efficiency limits as described in the previous sections.

\begin{figure}
    \centering
    \includegraphics[width=1\linewidth]{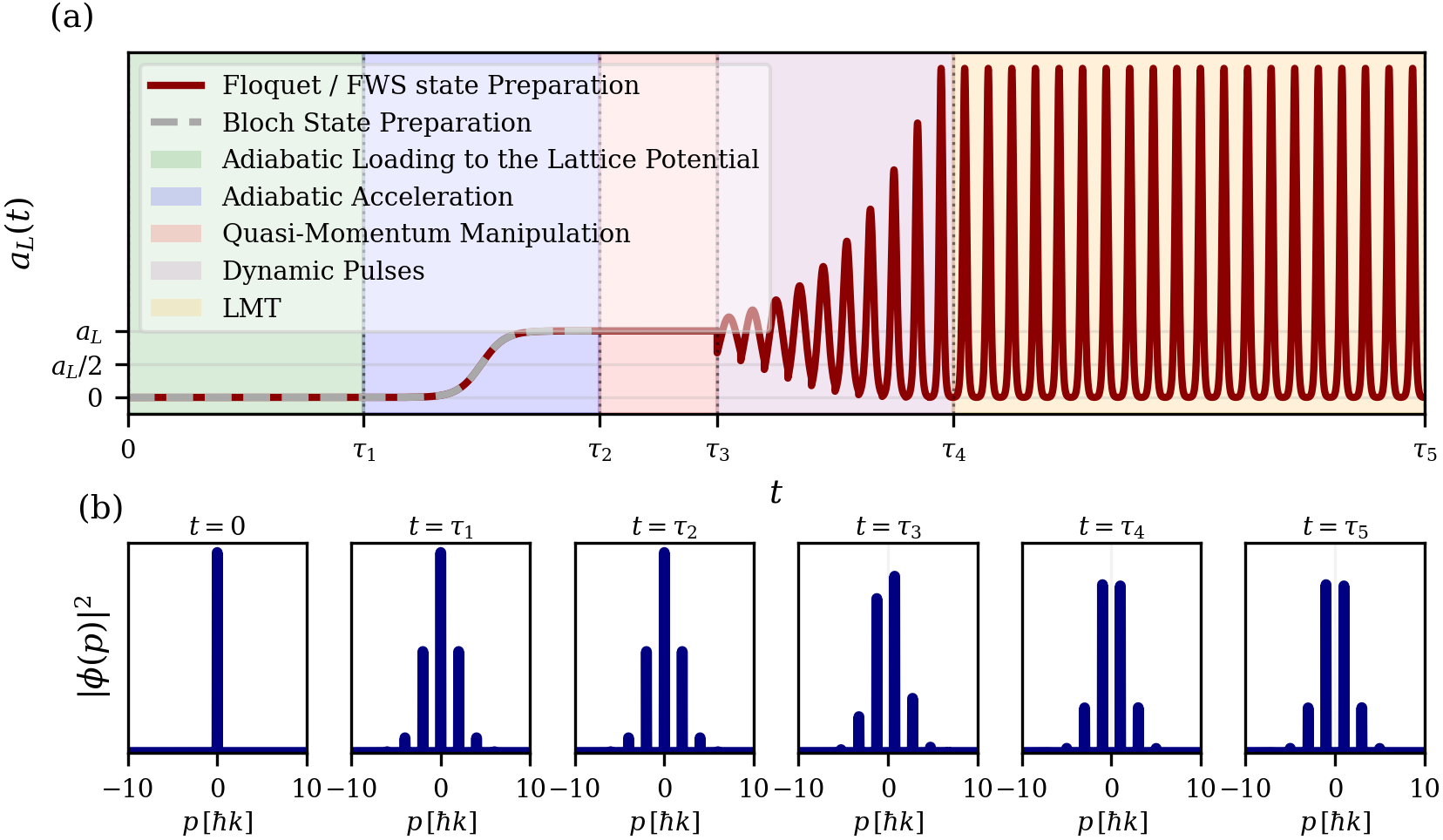}
    \caption{(a) Schematic illustration of lattice acceleration in arbitrary units versus time $t$ for adiabatic preparation of Floquet state $\ket{\phi_0(\kappa)}$ and SBDs LMT pulse. The pulse sequence is split into distinct steps: $t\in [0, \tau_1]$ adiabatic ramping of lattice potential; $t \in (\tau_1, \tau_2]$ adiabatic ramping of lattice acceleration; $t \in (\tau_2, \tau_3]$ BOs controlling the quasi-momentum $\kappa$; $t \in (\tau_3, \tau_4]$ adiabatic change of lattice acceleration function $\eta$; $t \in (\tau_4, \tau_5]$ SBDs LMT pulse. Note that the first two steps correspond to adiabatic Bloch state preparation~\cite{adiabatic-AI}. (b) Momentum representation probability amplitude of wavefunction in arbitrary units at time-steps $t=0, \tau_1, \tau_2, \tau_3, \tau_4, \tau_5$ in the co-moving lattice frame.}
    \label{fig:5}
\end{figure}

\textit{Adiabatic Preparation---}
To validate our theoretical framework, we perform numerical simulations using the ``Universal Atom Interferometer Simulator'' (UATIS)~\cite{UATIS}. First, we define an adiabatic preparation sequence, as demonstrated in Fig.~\ref{fig:5}. This preparation method represents a different, straightforward approach compared to the optimal control discussed in~\cite{Toulouse2} allowing us to control the quasi-momentum of the prepared state. Next, we numerically quantify the losses to the continuum by computing the decayed fraction of the Floquet states after a LMT sequence (c.f. Fig.~\ref{fig:2}\,(b,c), as well as the phase uncertainty arising from pulse-to-pulse fluctuations and lattice depth variations and find excellent agreement with our analytical model.

\textit{Conclusion---}
We have developed a unified theoretical framework for large-momentum-transfer methods in optical lattices based on a Floquet description of elastic light–atom scattering. Within this framework, we introduced a continuous parametrization that encompasses Bloch oscillations and sequential Bragg diffraction as limiting cases, enabling a direct and quantitative comparison of their efficiency and phase properties. We further proposed an adiabatic preparation scheme for Floquet states that provides controlled access to the relevant scattering regimes. The validity of our approach is established through quantitative agreement with exact numerical solutions of the Schrödinger equation as well as with recent state-of-the-art experimental results.
Beyond reproducing existing implementations, the framework identifies clear pathways for further optimization, including the realization of higher-efficiency pulses at shorter durations and the development of schemes that are more robust against dephasing. Such optimizations can be systematically explored both within the parameter space considered here and through alternative parameterizations of the underlying Hamiltonian. Taken together, the fundamental efficiency limits identified for experimentally relevant parameters and the demonstrated flexibility of the Floquet approach provide a solid foundation for the design of next-generation high-precision atom-interferometric experiments.

\textit{Acknowledgments---}
We thank Ernst Rasel, Pierre Cladé, and David Weld for fruitful discussions. A.A and P.M would like to express their gratitude to their colleagues in T-SQUAD and LCAR for their feedback and support. This work was funded by the Deutsche Forschungsgemeinschaft (German Research Foundation) under Germany’s Excellence Strategy (EXC-2123 QuantumFrontiers Grant No.
390837967) and through CRC 1227 (DQ-mat) within project No. A05. We also acknowledge funding by the by the AGAPES project – grant No.
530096754 within the ANR-DFG 2023 Programme.
\bibliographystyle{apsrev4-2}
\bibliography{references}
\end{document}


\title{Supplemental Material: Fundamental Limits of Large Momentum Transfer in Optical Lattices}

\author{Ashkan Alibabaei}
\thanks{These authors contributed equally to this work.}
\affiliation{Institute of Quantum Optic, Leibniz University of Hannover, Welfengarten 1
30167 Hannover, Germany}

\author{Patrik Mönkeberg}
\thanks{These authors contributed equally to this work.}
\affiliation{Institute of Quantum Optic, Leibniz University of Hannover, Welfengarten 1
30167 Hannover, Germany}
\affiliation{Institute of Theoretical Physics, Leibniz University of Hannover, 
Appelstrasse 2
30167 Hannover, Germany}
\affiliation{Institute for Theoretical Physics, University of Innsbruck, 6020 Innsbruck, Austria}

\author{Florian Fitzek}
\thanks{Currently at ASML Netherlands B.V., De Run 6501, 5504 DR Veldhoven.}
\affiliation{Institute of Quantum Optic, Leibniz University of Hannover, Welfengarten 1
30167 Hannover, Germany}
\affiliation{Institute of Theoretical Physics, Leibniz University of Hannover, 
Appelstrasse 2
30167 Hannover, Germany}

\author{Michael Werner}
\affiliation{Institute of Quantum Optic, Leibniz University of Hannover, Welfengarten 1
30167 Hannover, Germany}

\author{Alexandre Gauguet}
\affiliation{Laboratoire Collisions Agrégats Réactivité, FERMI, Université de Toulouse and CNRS UMR5589, 118 Route de Narbonne, F-31062 Toulouse, France}

\author{Baptiste Allard}
\affiliation{Laboratoire Collisions Agrégats Réactivité, FERMI, Université de Toulouse and CNRS UMR5589, 118 Route de Narbonne, F-31062 Toulouse, France}

\author{Klemens Hammerer}
\email[Contact author: ]{klemens.hammerer@uibk.ac.at}
\affiliation{Institute of Theoretical Physics, Leibniz University of Hannover, 
Appelstrasse 2
30167 Hannover, Germany}
\affiliation{Institute for Theoretical Physics, University of Innsbruck, 6020 Innsbruck, Austria}
\affiliation{Institute for Quantum Optics and Quantum Information of the Austrian Academy of Sciences, 6020 Innsbruck, Austria}

\author{Naceur Gaaloul}
\email[Contact author: ]{gaaloul@iqo.uni-hannover.de}
\affiliation{Institute of Quantum Optic, Leibniz University of Hannover, Welfengarten 1
30167 Hannover, Germany}

\date{\today}

\maketitle

\tableofcontents
\newpage
\section{Hamiltonian and lattice parameters}
In this section, we present an overview of the Hamiltonian and control parameters in three different reference frames and list their properties. Each of the presented reference frames is relevant for different parts of the discussion or calculation of the model presented in the main article. 
\subsection{Laboratory frame}
In the laboratory frame the Hamiltonian has the following form
    \begin{align}
H_\lab(t)=\frac{\hat{p}^2}{2m}+V(t)\cos^2[k_L(\hat{x}-x_L(t))]
    \end{align}
    with a periodically modulated lattice depth
    \begin{align}
        V(t)=V(t + T_F)
    \end{align}
and a lattice position $x_L(t)$ corresponding to a periodically modulated acceleration
    \begin{align}
        \begin{split}
            x_L(t) &= \int_{0}^{t} \dd{t'} \int_{0}^{t'}\dd{t''} a_L(t''),\\
            p_L(t) & = \int_{0}^{t}\dd{t'}m a_L(t'), \\
            a_L(t)&=a_L(t + T_F).
        \end{split}
    \end{align}
Note that the lattice depth is given by the effective two-photon Rabi frequency $\Omega(t)$ via $V(t) = 2\hbar \Omega(t)$. Additionally we define the acceleration averaged over one period as     
\begin{align}\label{eq:velocityincrem}
    \bar a_L&:=\frac{1}{T_F}\int_{t_0}^{t_0+T_F}dt\, a_L(t),
\end{align}

\subsection{Reduced Lattice Frame}\label{sup_sec:reduced_frame}
The reduced lattice frame is defined as a co-moving reference frame where the momentum is still measured with respect to the laboratory frame. The Hamiltonian takes the form 
\begin{align}
    H_\rlf(t)&=\frac{(\hat{p}-p_L(t))^2}{2m}+V(t)\cos^2(k_L\hat{x}).
\end{align}
We used $H_\rlf(t)=U(t)H_\lab U^\dagger(t)+\i\hbar\dot{U}U^\dagger(t)$ and
\begin{align}
    U(t)&=\exp\left(\frac{\i}{\hbar} \bigg(\hat{p}x_L(t) +\gamma_U(t)\bigg)\right),
\end{align}
where $\dot{\gamma}_U(t)=p_L(t)^2/2m$. We used the identities 
\begin{align}
    \e^{\hat{X}}\hat{Y}\e^{-\hat{X}} &= \sum_{n=0}^{\infty}\frac{[ \hat{X}, \hat{Y}]_m}{m!} = \hat{Y} + [\hat{X}, \hat{Y}] + \frac{1}{2!}[\hat{X}, [\hat{X}, \hat{Y}]] + \dots, \\
    \dv{t}\e^{\hat{X}(t)} &= \e^{\hat{X}(t)}\int_{0}^{1}\dd{\lambda}\e^{-\lambda \hat{X}(t)}\left(\dv{}{t}\hat{X}(t)\right)\e^{\lambda \hat{X}(t)}.
\end{align}
Note that the Hamiltonian in the reduced lattice frame $H_\rlf(t)$ has a discrete translation invariance with period $d=\pi/k_L$, that is
\begin{align}\label{eq:Hrlfperiod}
    \hat{T}H_\rlf(t) \hat{T}^\dagger&=H_\rlf(t),
\end{align}
where 
\begin{align}
    \hat{T}&=\e^{\i \hat{p} d/\hbar}
\end{align}
is the discrete spatial translation operator, shifting by distance $d$.

\subsection{Lattice Frame}
Finally in the lattice frame, which is co-moving in both position and momentum, after using the identity $\cos{2x}=2\cos^2{x}-1$, the Hamiltonian reads
\begin{align}\label{eq:H_lat}
    H_\lat(t)&=\frac{\hat{p}^2}{2m}+V(t)\cos(2k_L\hat{z})+ma_L(t)\hat{z}.
\end{align}
This follows from the unitary transformation $H_\lat(t)=\Tilde{U}(t)H_\rlf(t) \Tilde{U}^\dagger(t)+\i\hbar\dot{\Tilde{U}}(t)\Tilde{U}^\dagger(t)$ with
\begin{align}\label{eq:V}
    \tilde{U}(t)&=\exp\bigg(-\i \hat{x} p_L(t)/\hbar\bigg).
\end{align}
The Hamiltonian in the lattice frame is time-periodic but \textit{not} translation invariant
\begin{align}\label{eq:H_invariances}
    H_\lat(t+T_F)&=H_\lat(t), & \hat{T}H_\lat(t) \hat{T}^\dagger&= H_\lat(t)+ma_L(t)d.
\end{align}

Note that the general Hamiltonian in Eq.~\eqref{eq:H_lat} describes both Bloch Oscillations (BOs) and Sequential Bragg Diffractions (SBDs). For BOs we have a constant acceleration $a_L(t)\equiv a_0$ and a constant lattice depth $V(t)\equiv V_0$, while for SBDs we have a $\delta$-comb acceleration  $a_L(t) = a_0\sum_n\delta(t-nT_F)$ and a constant lattice depth.

\section{Floquet solution}
The time-periodicity of the Hamiltonian in the lattice frame in Eq.~\eqref{eq:H_invariances} enables us to find the eigenstates and eigenenergies of the system via Floquet theory.   
\subsection{Floquet basis}
Despite the fact that $H_\text{lat}(t)$ has no spatial periodicity in Eq.~\eqref{eq:H_invariances}, the \emph{Floquet operator} $U_\text{lat}(t_0 + T_F, t_0)$ is invariant under discrete translations by $d$, provided the momentum $\Delta p_L=p_L(t_0+T_F)-p_L(t_0)$ imparted per period is a multiple of $2\hbar k_L$. We show this by expressing the time evolution operator in the lattice frame as
\begin{align}
    U_\lat(t,t_0)=\Tilde{U}(t)U_\rlf(t,t_0)\Tilde{U}^\dagger(t_0),
\end{align}
where $\Tilde{U}(t)$ is the unitary given in Eq.~\eqref{eq:V} and $U_\text{rlf}(t, t_0)$ is the time evolution operator in the reduced lattice frame. The latter is spatially periodic $\hat{T}U_\text{rlf}(t, t_0)\hat{T}^\dagger = U_\text{rlf}(t, t_0)$ due to the spatial periodicity of the Hamiltonian in the reduced lattice frame in Eq.~\eqref{eq:Hrlfperiod}. In the lattice frame we then have 
\begin{align}
    \hat{T}U_\lat(t,t_0)\hat{T}^\dagger = \e^{-\i d(p_L(t)-p_L(t_0))/\hbar}U_\lat(t,t_0),
\end{align}
and for the Floquet operator we can conclude 
\begin{align}
    \hat{T}U_\lat(t_0+T,t_0)\hat{T}^\dagger=\e^{-\i d \Delta p_L/\hbar}U_\lat(t_0+T,t_0).
\end{align}
Thus, spatial periodicity requires $d\Delta p_L/\hbar = 2\pi n$, which according to Eq.~\eqref{eq:velocityincrem} and $d=\pi/k_L$ is equivalent to 
\begin{align}\label{eq:cond}
    \bar{a}_L = \frac{2n\hbar k_L}{mT_F}.
\end{align}
Under condition in Eq.~\eqref{eq:cond} we thus have for all $t_0$
\begin{align}\label{eq:spaceinvariance}
    [\hat{T},U_\lat(t_0+T_F,t_0)]=0.
\end{align}
Note that for the special case of BOs without an externally imposed time period $T_F$, the condition in Eq.~\eqref{eq:cond} sets the connection between the constant acceleration $a_L\equiv a_0$ and the period. The case $n=1$ sets the shortest period, the Bloch period $T_B = 2\hbar k_L / m a_0$. For any other pulse sequence, with a chosen acceleration $a_L(t)$ and period $T_F$, the condition in Eq.~\eqref{eq:cond} determines the number $n$ of momentum quanta $2\hbar k_L$ imparted per period. If we define the equivalent Bloch period for a given average acceleration $\bar{a}_L$ via $T_B = 2\hbar k_L/m \bar{a}_L$, then $n=T_F/T_B$ gives the ratio of the external time period to the equivalent Bloch period. This allows us to directly compare the case of BOs to any other case with constant lattice depth. 

The fact that the Floquet operator is invariant under discrete translations in Eq.~\eqref{eq:spaceinvariance} implies that there exists a basis of common eigenstates, the Floquet states
\begin{align}
U_\lat(t_0+T_F,t_0)\ket{\phi_\alpha(\kappa,t_0)}&=\e^{-\i \mathcal{E}_\alpha(\kappa,t_0)T_F/\hbar}\ket{\phi_\alpha(\kappa,t_0)} \\
    \hat{T}\ket{\phi_\alpha(\kappa,t_0)}&=\e^{\i\kappa d}\ket{\phi_\alpha(\kappa,t_0)}\label{eq:No1}
\end{align}
with quasi-momentum $-k_L\leq\kappa < k_L$ and a band index $\alpha$ \footnote{Note that without loss of generality, we can set $t_0=0$. For the sake of consistency we keep track of $t_0$ in the supplementary material, whereas in the main text we assume the initial time to be zero.}. Due to the discrete spatial invariance of the Floquet operator (cf. Eq.~\eqref{eq:spaceinvariance}), the quantum number $\ell$ obtained from reconstructing the modulo $2\pi$ invariance of the quasi-energies can be identified with the lattice site.

The time-evolved Floquet states are defined as 
\begin{align}
    \ket{\Phi_\alpha(\kappa_0,t)}:=\e^{\i \mathcal{E}_{\alpha, \ell}(\kappa)(t -t_0)/\hbar}U_\lat(t,t_0)\ket{\phi_\alpha(\kappa_0,t_0)}
\end{align}
where the dynamical phase $\e^{\i \mathcal{E}_{\alpha, \ell}(\kappa)(t -t_0)/\hbar}$ fixes to the $U(1)$ gauge freedom of the Floquet states and ensures that the Floquet states are periodic in time.

\subsection{Quasi-momentum}
In the general case, the Floquet states as well as the quasi-energy spectrum depend on the quasi-momentum $\kappa$. Note that in the case of constant acceleration and lattice depth, i.e. for BOs, it can be shown that the spectrum is independent of the quasi-momentum. Subsequently a basis change to Wannier-Stark states can be performed, where the states are likewise independent of $\kappa$~\cite{Gluck}. However, since the quasi-momentum $\kappa$ in general remains a relevant quantum number, we will solely use the Floquet basis here. 

Therefore for a given lattice depth $V_0$ and Floquet period $T_F$, two degrees of freedom remain: (i) the acceleration type that determines the LMT method, i.e. $\eta$ in the parametrization which we presented in the main text, and (ii) the quasi-momentum $\kappa$ which is determined by the relative momentum between the atoms and the lattice. For an atomic cloud with a given temperature, multiple quasi-momentum states will contribute to the overall LMT sequence. Due to the periodic behavior of both the quasi-energy spectrum and Floquet states with the quasi-momentum, it suffices to explore only the first Brillouin zone, $\kappa \in [-k_L, k_L)$, to identify the optimal quasi-momentum for implementing LMT sequences at given lattice parameters. 

\begin{figure}
    \centering
    \includegraphics[width=0.5\linewidth]{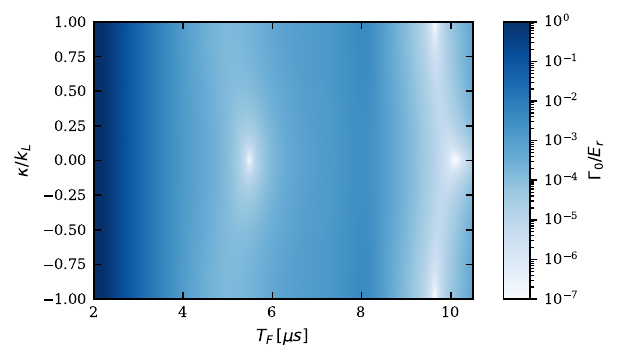}
    \caption{Tunneling loss rate $\Gamma_0$ of Floquet groundstate in the limit of SBDs ($\eta =100$) versus Floquet period $T_F$ and quasi-momentum $\kappa$ for a lattice depth $V_0=50E_r$. The colorbar is cut off at $10^{-7}$ to increase contrast.}
    \label{fig:kappa_dependency}
\end{figure}

Fig.~\ref{fig:kappa_dependency} shows the quasi-momentum dependency across one Brillouin zone of the tunneling loss rate of the fundamental Floquet band $\Gamma_0(\kappa)$ for different Floquet periods. This covers both cases studied in the main text, namely $\kappa =0$ and $\kappa=k_L$, where the latter is experimentally explored by Rodzinka et al.~\cite{Toulouse2}. Fig.~\ref{fig:kappa_dependency} especially shows that for the assumed lattice depth of $V_0=50E_r$, $\kappa=0$ is the optimal choice for efficient LMT at high transfer rates in the SBDs regime. Additionally, the anti-resonances have a finite width in $\kappa$, which allows for utilization of the corresponding efficiency enhancement even for finite-temperature initial atomic clouds, given that the temperature is below a certain threshold. \\
Due to the deep lattice depths we consider here, the real Floquet bands $E_\alpha(\kappa)$ are flat, more precisely the relative band width is on the order of $\Delta E_\alpha / E_\alpha \sim 10^{-5}-10^{-7}$ for the parameter regimes we consider here.

\section{Numerical methods}
In this section, we present the numerical methods used for calculating the Floquet spectrum in the general case, as well as the method used for sorting said spectrum.
\subsection{Diagonalization of Floquet operator}
We now present a numerical routine to calculate the Floquet states and the complex spectrum. We work in the plane-wave momentum basis $\ket{2n\hbar k_L + \hbar \kappa}$ with $n \in \mathbb{Z}$ and quasi-momentum $\kappa \in [-k_L, k_L)$. To avoid representing the position operator $\hat{x}$ in this basis, we work in the reduced lattice frame (cf. \ref{sup_sec:reduced_frame}) and rewrite the Floquet operator in the lattice frame as 
\begin{align}\label{eq:floquet_op}
    U_\text{lat}(t_0+T_F, t_0) = \e^{-\i 2k_L \hat{x}}U_\text{rlf}(t_0 + T_F, t_0),
\end{align}
where 
\begin{align}
    U_\text{rlf}(t_0 + T_F, t_0) = \mathcal{T}\exp(-\frac{\i}{\hbar} \int_{t_0}^{t_0+T_F}\dd{t} H_\text{rlf}(t)),
\end{align}
and $\mathcal{T}$ denotes time-ordering. To compute $U_\text{rlf}(t_0 + T_F, t_0)$, we define a discrete time-grid with $J$ equidistant grid points $t_j = (j - 1/2)\dd{t}$ for $j \in \{1, \dots, J\}$ and $\dd{t} = T_F / J$. For sufficiently small time-steps, we approximate 
\begin{align}
    U_\text{rlf}(T)\simeq \prod_{j=1}^{J}\exp(-\i H_\text{rlf}(t_j)\dd{t}/\hbar).
\end{align}
Next, we expand the Hamiltonian in the momentum basis $\ket{2n\hbar k_L + \hbar \kappa}$
\begin{align}\label{eq:reduced_hamiltonian}
    H_\rlf(t_j) = \sum_{n\in \mathbb{Z}}\left( \left(\frac{(2n\hbar k_L + \hbar \kappa - p_L(t_j))^2}{2m} + \frac{V_0}{2}\right)\ketbra{2n\hbar k_L + \hbar \kappa}{2n\hbar k_L + \hbar \kappa} + \frac{V_0}{4}\left(\ketbra{2(n+1)\hbar k_L+ \hbar \kappa}{2n\hbar k_L + \hbar \kappa} + \text{h.c.} \right)\right).
\end{align}
Note that the average AC Stark shift $V_0/2$ can be subtracted here for the computation of linewidths and states, or for the energies with AC stark shift suppression. To fully expand the Floquet operator in Eq.~\eqref{eq:floquet_op} in the momentum basis, we now consider the momentum shift operator $\e^{-\i 2k_L \hat{x}}$, which takes the form 
\begin{align}\label{eq:momentum_shift}
    \e^{-\i 2k_L \hat{x}} = \sum_{n \in \mathbb{Z}} \ketbra{2(n-1)\hbar k_L + \hbar \kappa}{2n\hbar k_L + \hbar \kappa}.
\end{align}
The discretized Floquet operator can then be obtained by combining Eqs.~\eqref{eq:floquet_op}, \eqref{eq:reduced_hamiltonian} and \eqref{eq:momentum_shift}. Note that we so far left the quasi-momentum undetermined, however the discretized Floquet operator belongs to the subspace of a specific quasi-momentum $\kappa$. The discretizations is based on a time grid with $J$ grid points and a momentum grid truncated at $n \in [-N, N]$ with $2N+1$ grid points. This leads to a low-dimensional matrix, that can be efficiently constructed and diagonalized using standard libraries for scientific computing, however convergence in the discretizations parameters $J$ and $N$ has to be ensured for accurate results. Truncating the momentum shift operator in Eq.~\eqref{eq:momentum_shift} leads to a non-unitary matrix and subsequently to complex eigenvalues of the discretized Floquet operator $U_\lat(t_0 + T_F, t_0)$ of the form $\exp(-\i \mathcal{E}_nT_F/\hbar) = \exp(-\i (E_n - \i \Gamma_n/2)T_F/\hbar)$. From these eigenvalues we can determine the quasi-energies $\mathcal{E}_n$ modulo $2\pi\hbar/T_F$. To compute the Floquet-Wannier-Stark energies $\mathcal{E}_{\alpha, \ell}$, the modulo operation needs to be reconstructed and furthermore the momentum state indices $n$ need to be correctly mapped onto the band indices $\alpha$.

\begin{figure}
    \centering
    \includegraphics[width=0.5\linewidth]{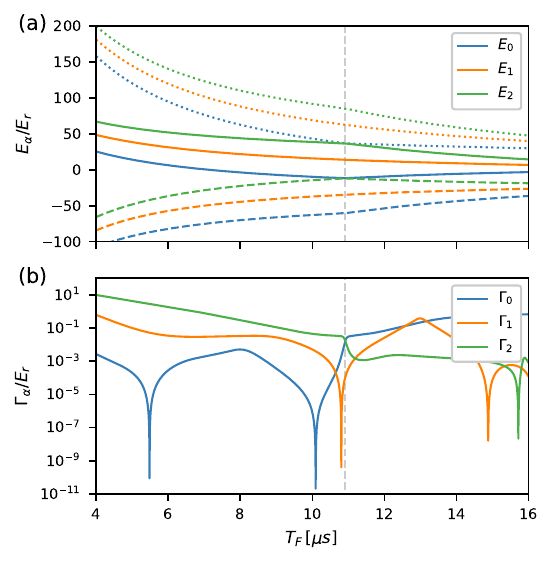}
    \caption{(a) Quasi-energies versus Floquet period at lattice depth $V_0=50\,E_r$. The solid lines correspond to lattice site $\ell=0$, the dashed lines to $\ell = -1$ and the dotted lines to $\ell=+1$. The vertical dashed line marks the Floquet period where the energies of the lowest and the second excited Floquet bands show an avoided crossing. (b) First three linewidths versus Floquet period at lattice depth $V_0 = 50\,E_r$. Vertical dashed line marks the Floquet period where the linewidths of lowest and second excited Floquet bands show a real crossing, exactly where the real energies show an avoided crossing.}
    \label{fig:extended_spectrum}
\end{figure}

\subsection{Quasi-energy sorting}
Finding the correct quasi-energy band sorting in Floquet systems is a common problem, since the band gaps are usually larger than the modulo operation, up to which one can obtain the quasi-energies. Hence sorting according to increasing energies $E_n$ here will not result in the correct identification of $n$ and $\alpha$. In the Wannier-Stark description of Bloch oscillation LMT pulses, it has been proven effective to sort according to increasing linewidths \cite{Bloch_paper}, which are usually much smaller than the remainder $\text{Im}(\mathcal{E}_{\alpha, 0}) = \Gamma_\alpha / 2 \ll 2\pi \hbar /T_F$. However this kind of sorting will fail real crossings of the linewidths of different Floquet bands and create artificial avoided crossings, which in the generalized case happens regularly. Nonetheless, in the limit of small Floquet periods $T$, sorting by increasing linewidths will give an unambiguous and correct mapping between the momentum state indices $n$ and band indices $\alpha$. To resolve the issue of artificial avoided crossings we then impose an adiabaticity condition on the Floquet states of Systems with varied Floquet period $T_F$ or lattice depth $V_0$
\begin{align}\label{eq:sorting_cond}
    \braket{\phi_\alpha[\lambda]}{\phi_\beta[\lambda + \dd{\lambda}]}\simeq \delta_{\alpha \beta} 
\end{align}
for $\dd{\lambda}$ sufficiently small and  $\lambda = T_F$ or $\lambda=V_0$. Iteratively sorting by the condition in Eq.~\eqref{eq:sorting_cond} on top of initially sorting by increasing linewidths results in the correct mapping between $n$ and $\alpha$ for all Floquet periods $T_F$ and lattice depths $V_0$. It is important to note that this sorting method can cause non-holonomic behavior~\cite{Anholonomies-Floquet}. Depending on which parameter is adiabatically varied and sorted by, the resulting state can be different. More precisely, the path taken in parameter space determines the resulting quantum state. Moreover, not only the eigenstate differs, as can for example be observed in the field of geometric phases, but also so does the eigenenergy. Therefore taking different paths in parameter space can result in occupying completely different eigenspaces, which is known as Cheon's anholonomies~\cite{cheon, Anholonomies-Floquet}. This can be relevant for adiabatic preparation of the Floquet states, where multiple parameters are varied adiabatically in a specific order which corresponds to a specific path in parameter space.

\section{Spontaneous emission losses}
\begin{figure}
    \centering
    \includegraphics[width=0.4\linewidth]{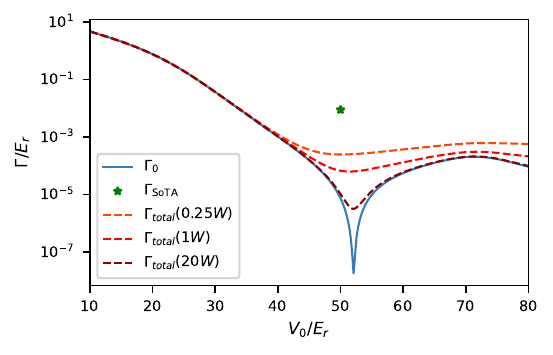}
    \caption{Fundamental tunneling loss rate and total loss rate including fundamental tunneling losses and spontaneous emission losses for a Floquet period $T_F=5.3\,\mu \mathrm{s}$ versus lattice depth $V_0$. The spontaneous emission losses are estimated in the presented naive model via Eq.~\eqref{eq:gamma_sp_final} for three different laser powers $P=250\, \mathrm{mW}, 1\, \mathrm{W}, 20\,\mathrm{W}$ and a fixed beam waist of $w=1.6\,\mathrm{mm}$. Additionally, the green point shows the loss rate reported in~\cite{Toulouse2} for comparison.}
    \label{fig:spontaneous_emission}
\end{figure}

In this section, we estimate the spontaneous emission losses for LMT pulses that use a constant lattice depth. To simplify the analysis, we consider all atoms that undergo a spontaneous scattering event as lost and therefore reducing the overall number of atoms. In general these atoms can undergo transitions to different Floquet bands, resulting in more complex dynamics. However, for sufficiently small spontaneous emission rates, these effects can be neglected. We estimate the spontaneous scattering rate following~\cite{spontaneous_emission} as
\begin{align}
    \hbar \Gamma_\text{sp} = \frac{\Gamma_\text{nat}}{\vert \Delta \vert}\langle V(\hat x)\rangle,
\end{align}
where $\Gamma_\text{nat}$ is the natural linewidth of the relevant atomic transition, $\Delta$ is the detuning and $\langle \cdot \rangle$ denotes the average with respect to the atomic state. Since we only consider deep lattices in this work, we can apply the harmonic approximation to the average potential of a blue-detuned lattice
\begin{align}\label{eq:gamma_sp}
    \hbar \Gamma_\text{sp} = \frac{\Gamma_\text{nat}}{\vert \Delta \vert} \frac{\sqrt{V_0 E_r}}{2}.
\end{align}
Next, we relate the lattice depth $V_0$ to the power $P$ and waist $w$ of a Gaussian laser beam via 
\begin{align}
    V_0 = \frac{3  c^2}{\omega_0^2}\frac{\Gamma_\text{nat}}{\vert \Delta \vert}\frac{ P}{ w^2},
\end{align}
where $\omega_0$ denotes the atomic resonance frequency and $c$ is the speed of light. Inserting this into Eq.~\eqref{eq:gamma_sp} results in
\begin{align}\label{eq:gamma_sp_final}
    \hbar \Gamma_\text{sp} = \frac{\omega_0^3 w^2}{6c^2P}\sqrt{V_0^3E_r} = \frac{2 \omega_0^3}{3\pi c^2}\frac{\sqrt{V_0^3 E_r}}{2 I_0},
\end{align}
where $I_0 = 2P / \pi w^2$ is the intensity. The spontaneous emission losses can the be calculated via 
\begin{align}\label{eq:spnt_losses}
    1 - P_\text{sp} = 1 - e^{-\Gamma_{sp} \tau},
\end{align}
where $\tau$ is the total time of the LMT sequence. In Fig.~\ref{fig:spontaneous_emission} we compare the tunneling losses, presented in the main article, and the spontaneous emission losses estimated via Eqs.~\eqref{eq:gamma_sp_final} and \eqref{eq:spnt_losses}. Note that the here presented model for spontaneous emission losses is quite simplified and for a more accurate description of precision measurement setups additional factors have to be taken into account. These include for example contributions from retro-reflected lattices, as well as the more complex dynamics due to spontaneous scattering events. However we stress that the specific model used here is not central to our conclusions, as it does not affect the results qualitatively. 

\section{Phase Uncertainty}
\begin{figure}
    \centering
    \includegraphics[width=0.5\linewidth]{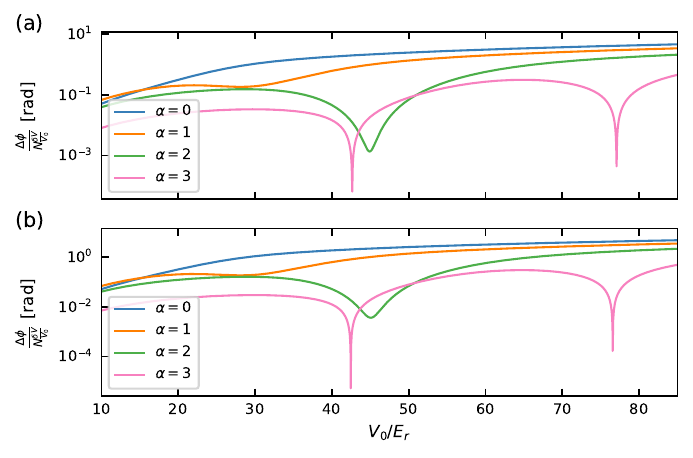}
    \caption{Phase uncertainty induced by lattice depth differences versus lattice depth $V_0$ for a Floquet period of $T_F=5.3 \mu s$. The specific Floquet period was chosen to cover the experiment~\cite{Toulouse2} at a lattice depth of $V_0 = 50E_r$ in the limit of SBDs. The phase uncertainty is determined for the first five Floquet bands in the limit of BOs $\eta=0.1$ (a) and the limit of SBDs $\eta=100$ (b). The average AC Stark shift is suppressed here.}
    \label{fig:4}
\end{figure}

One of the main contributors to the overall phase uncertainty is induced by intensity asymmetries of the lattice between the interferometer arms and noise that could couple to them. The phase imprinted by a single LMT pulse of $2\hbar k_L$ is given by $\phi_{\alpha} = E_{\alpha}T_F/\hbar$ for $\ell=0$. A lattice depth difference between the arms as well as pulse-to-pulse fluctuations within one arm, here both denoted by $\Delta V$, will therefore cause a phase shift of 
\begin{align}\label{eq:dephasing}
\frac{\Delta \phi_\alpha}{N \frac{\Delta V}{V_0}}=2\pi \biggl\vert\frac{\partial E_{\alpha}}{\partial V_0}\biggr\vert \frac{V_0}{ma_Ld}.
\end{align}
We normalize the phase shift to the number of Floquet periods $N$ and the relative lattice depth difference $\Delta V / V_0$. In Fig.~\ref{fig:4}\,(a,b) we show these phase changes for an exemplary Floquet period in the limit of BOs and of SBDs, respectively. As Fig.~\ref{fig:4} demonstrates, there is no qualitative difference between the phase uncertainty induced by BOs and SBDs and therefore the energies $E_{\alpha, \ell}$ only depend on the average acceleration, but not its temporal profile. As highlighted in the main text, there are particular lattice depths where the phase uncertainty $\Delta\phi_\alpha$ is zero for some excited states. At these parameters, the system is robust to dephasing caused by noisy lattice depth differences.

We can simulate the phase uncertainty predicted by Eq.~\eqref{eq:dephasing}, using the ``Universal atom interferometer simulator''~\cite{UATIS}, by defining two LMT sequences corresponding to the two arms of the interferometer based on our adiabatic preparation discussed in the last section of the main text. One arm evolves under the standard LMT sequence with an unperturbed lattice potential $V_0$, while for the other arm, a fluctuation $\Delta V$ is introduced to the LMT sequence. We obtain the dephasing from the relative phase accumulated between the two states, which is in excellent agreement with our analytical results.

\section{Sensitivity of multi-loop LMT interferometers}\label{Sec: Phase Analytical}
The realization of interferometers with large LMT can be challenging. As presented in the main text, the fundamental efficiency limits for SBDs based LMT methods permit sensitivity improvements up to $N_\text{LMT} \gtrsim 10^6$. In a conventional Mach-Zehnder interferometer such high LMT orders would correspond to thousands of kilometers of arm separation and are therefore unrealistic. In order to overcome this problem, we consider folded interferometer geometries to showcase the achievable sensitivity when performing the LMT pulses at their fundamental efficiency limits. We consider interferometers consisting of two beam splitter pulses and $2 N_\text{loop}$ mirror pulses in between, where $N_\text{loop} \in \mathbb{N}$. 

\subsection{Multi-loop interferometer phase}\label{sec:ml_phase}
We follow here the method presented in~\cite{phase_calculation} for the phase calculation. Under ideal conditions, the total interferometer phase for such geometries consists of the propagation phase $\Delta\Phi_\text{prop}$ and the light induced phase $\Delta\Phi_\text{lp}$
\begin{align}
    \Delta\Phi = \Delta\Phi_\text{prop} + \Delta\Phi_\text{lp}.
\end{align}
The propagation phase is given as 
\begin{align}
    \Delta\Phi_\text{prop} = \frac{1}{\hbar}\int [\mathcal{L}(z_u(t), \dot z_u(t)) - \mathcal{L}(z_\ell(t), \dot z_\ell(t))] \dd t
\end{align}
where $z_u$ and $z_\ell$ denote the upper and lower path respectively and with the Lagrangian 
\begin{align}
    \mathcal{L}(z, \dot z) = \frac{1}{2}m \dot z^2 - mgz + \frac{1}{2}m\gamma z^2.
\end{align}
Here $\gamma = \frac{\partial^2 \phi(z)}{\partial z^2}$ denotes the gravity gradient for the gravitational potential $\phi(z)$. Note that in the following we are using the ``unperturbed'' trajectories for the upper and lower path, i.e. the solutions to the Euler-Lagrange equations governed by $\mathcal{L}_0(z, \dot z) = \frac{1}{2}m \dot z^2 - mgz$. Additionally, we only consider Bloch-type acceleration for the LMT sequences here. This makes the calculations considerably easier and allows us to derive an analytical expression for the interferometer phase. Later on we will compare our analytical result to numerical simulations based on~\cite{phase_code}, and test the validity of the approximations made here.

We consider even number of loops, such that the upper and lower arm exchange roles between successive loops. Therefore the spacetime area of such two loops together is zero in total, and the interferometer is insensitive to constant accelerations. To show this on a more formal level, the propagation phase can be split into three contributions
\begin{align}\label{eq:prop_phase}
    \Delta\Phi_\text{prop} = \Delta\Phi_\text{kin} + \Delta\Phi_g + \Delta\Phi_\gamma
\end{align}
and we define the intrinsic interferometer time (time of a full LMT sequence) as 
\begin{align}
    T:= 2N_\mathrm{LMT}^\mathrm{loop}T_F + T_\mathrm{free},
\end{align}
where $N_\text{LMT}^\text{loop}$ again denotes the momentum transfer per loop and $T_\mathrm{free}$ the free evolution time between acceleration and deceleration of the LMT sequence. The three contributions to the propagation phase in Eq.~\eqref{eq:prop_phase} can then be written as 
\begin{align}
    \Delta\Phi_\text{kin} &= \frac{m}{2\hbar}\int[\dot z_u(t)^2 - \dot z_\ell(t)^2] \dd t  = 0 \\
    \Delta\Phi_g &= \frac{mg}{\hbar}\int [z_u(t) - z_\ell(t)]\dd t = 0 \\
    \Delta\Phi_\gamma &= \frac{m\gamma}{2\hbar} \int [z_u(t)^2 - z_\ell(t)^2]\dd t \nonumber \\
    &= 2\gamma N_\text{loop}\biggr( -\frac{\hbar k_L^2}{2m}T((2 + N^\text{loop}_\text{LMT})T + N^\text{loop}_\text{LMT}T_\text{free})^2 + k_LT^2((2 + N^\text{loop}_\text{LMT})T + N^\text{loop}_\text{LMT}T_\text{free})(2gT - v_0) \biggr) \label{eq:prop_phase_gamma},
\end{align}
where $v_0$ is the initial velocity of the atoms prior to the first beam splitter, and hence identical for both arms. Note that this expression only holds, if $v_0$ is the same for every two loops, i.e. we assume that after every second loop, or a time of $4T$, the whole sequence repeats itself perfectly. Otherwise a non-linear scaling with $N_\text{loop}$ and $v_0$ will be introduced. In the here employed first order perturbation theory approach, the light-induced phase is also found to vanish as the unperturbed trajectories lead to a perfectly symmetric interferometer sequence
\begin{align}
    \Delta\Phi_\text{lp} = 0.
\end{align}
Therefore the total interferometer phase is given by the gravity gradient contribution to the propagation phase 
\begin{align}
    \Delta\Phi = \Delta\Phi_\gamma,
\end{align}
which leads to a sensitivity to the gravity gradient at shot-noise limit of 
\begin{align}
    \Delta\gamma = \frac{1}{\sqrt{N_\text{atoms}}N_\text{loop}\mathcal{Q}},
\end{align}
where we introduced the interferometer geometry specific scale factor
\begin{align}
    \mathcal{Q}:=2k_LT^2\left((2 + N^\text{loop}_\text{LMT})T + N^\text{loop}_\text{LMT}T_\text{free}\right)(2gT - v_0) -\frac{\hbar k_L^2}{m}T\left((2 + N^\text{loop}_\text{LMT})T + N^\text{loop}_\text{LMT}T_\text{free}\right)^2.
\end{align}
\subsection{Relaunch scheme}
We consider a specific relaunching scheme for the atoms to make sure the interferometers average  position remains constant. We consider relaunching both interferometer arms simultaneously every second loop. To this end the atoms are initially launched with velocity $v_0$ and relaunched with velocity $v_\text{relaunch}$, such that the atoms again have a velocity of $v_0$ after every relaunch. This matches the condition we previously introduced in Sec.~\ref{sec:ml_phase} for the calculation of the propagation phase (cf. Eq.~\eqref{eq:prop_phase_gamma}). Note that while the interferometer arms are spatially overlapping at the relaunch points, they are at all times separated in momentum space. 

The corresponding required initial velocity for each $4T$-sequence is given by 
\begin{align} \label{eq:init_velo}
    v_0 = 2 g T - N^\text{loop}_\text{LMT} \frac{\hbar k_\text{L}}{m} \frac{N^\text{loop}_\text{LMT} T_\text{F} + T_\text{free}}{T},
\end{align}
and the atoms are relaunched with $v_\text{relaunch}=4gT$. At each relaunch point we apply the same amount of LMT pulses simultaneously to both arms to accelerate both arms again to a velocity of $v_0$. For the sensitivity of the whole interferometer sequence, we also take the losses of the relaunches into account, assuming the same efficiency for the relaunch-LMT pulses as for the regular LMT pulses.

\begin{figure}
    \centering
    \includegraphics[width=0.5\linewidth]{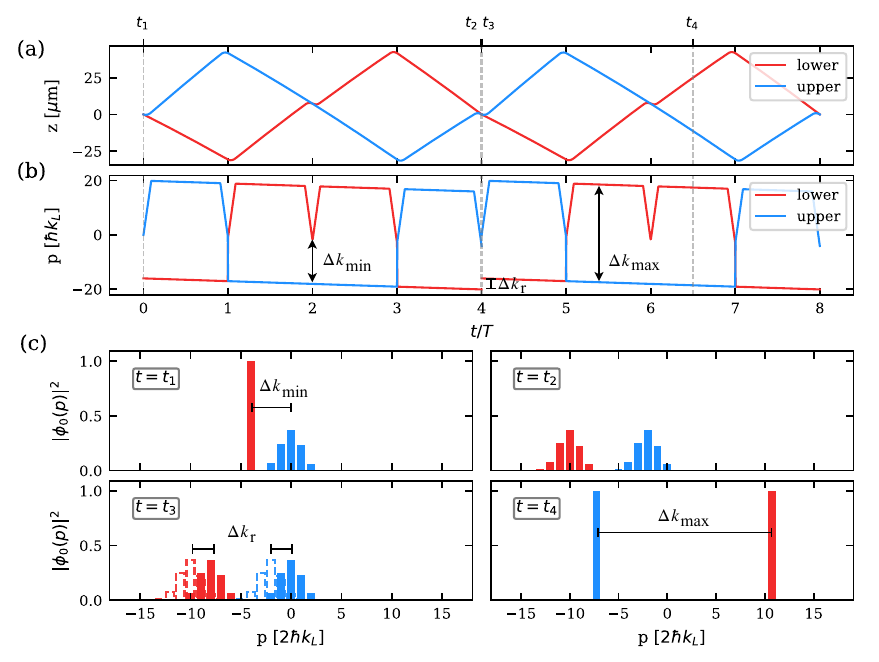}
    \caption{(a) Space-time diagram for 4-loop sequence with $N_\mathrm{LMT}^\mathrm{loop}=10$, $T_\mathrm{free}=0.005\,\mathrm{s}$ and fourth order double-Bragg diffraction beam splitters and mirrors. 
    (b) Corresponding momentum-time diagram for the 4-loop sequence. The minimum separation in momentum space $\Delta k_\mathrm{min}$ is set by the beam splitters and mirrors, while the maximum momentum separation $\Delta k_\mathrm{max}$ additionally depends on the LMT per loop $N_\mathrm{LMT}^\mathrm{loop}$. The arms switch roles in momentum space at the mirrors at times $t_M = T, 3T, 5T, 7T$; otherwise they are always separated in momentum space by $\Delta k_\mathrm{min}$ including at the relaunch time $t = 4T$. The momentum jump of both arms by $\Delta k_r = mv_\mathrm{relaunch}$ at $t = 4T$ indicates the relaunch.
    (c) Atomic states of the upper (blue) and lower (red) interferometer arm in momentum representation at four different times throughout the 4-loop sequence. In the optical lattice the atoms are in Floquet states, which spread out over multiple momentum bins. In free space the atoms are here assumed to be in a single momentum state. $t_1 = 0$ denotes the beginning of the sequence immediately after beam splitter pulse is applied; $t_2 = (4-\epsilon)T$, with $\epsilon \ll 1$, denotes the instant immediately before relaunch; and $t_3 = (4 + \epsilon)T$ denotes the time directly after relaunch. Dashed bars indicate the states prior to relaunch. Finally, $t_4 = 6T + N_\mathrm{LMT}^\mathrm{loop}T_F + T_\mathrm{free}/2$ denotes a time during free evolution when the interferometer arms are at large momentum separation within an LMT sequence.}
    \label{fig:mom-time-diag}
\end{figure}

To avoid overlap between the Floquet states of the different interferometer arms during relaunch, the momentum separation between them at the relaunches has to be large enough. To ensure this, high-order double-Bragg diffraction beam splitter and mirror pulses can be used similar to the experiment of reference~\cite{Gebbe}. 
Fig.~\ref{fig:mom-time-diag}(a, b) show the trajectories and momenta of the interferometer arms for an example sequence of four loops, illustrating the required minimal momentum separation and the relaunch. Fig.~\ref{fig:mom-time-diag}(c) demonstrates this on the level of the atomic wavefunctions of the two interferometer arms throughout the sequence. The relaunch stage could be achieved by two optical lattices with orthogonal polarizations sufficiently separated in momentum space to suppress crosstalk between the two arms, each resonant with solely one of the optical lattices.

Note that different beam splitters and mirrors affect the classical trajectories and therefore the propagation phase. However we find that for the parameters we consider in the main text, namely $T_\text{free}=0.3\,\mathrm{s}$ and a maximum of $N^\text{loop}_\text{LMT}=1000$, this difference is minimal. Especially for a momentum transfer per loop above $N^\text{loop}_\text{LMT}=100$, the impact of different beam splitter and mirror pulses is negligible.

\subsection{Comparison to numerical simulations with SBDs}

As mentioned in Sec.~\ref{sec:ml_phase}, for the sake of simplicity we considered Bloch-type acceleration for the LMT sequence in order to derive an analytical expression for the phase \footnote{Note that one could also calculate the total phase shift, which includes the contribution from the laser phase and the separation phase, all evaluated along the perturbed trajectories. However, such an approach would lead to an increasingly large separation at the output port, which would manifest itself as a phase contribution, but ultimately degrade the contrast. In practice, it is essential for all interferometer sequences to implement a gravity gradient mitigation scheme; therefore, we will not delve into the details here and analyze all phases using the first order perturbed propagation phase.}. In order to justify our result we compare it to numerical simulations where we implement the Bragg-type accelerations, i.e. a series of instantaneous momentum kicks with strength $2\hbar k_\text{L}$.

Recall that the analytical phase from Eq.~\eqref{eq:prop_phase_gamma} can be rewritten as
\begin{align}
    \Delta \Phi_\text{Analytical} &= 2 \gamma N_\text{loop} k_\text{L} T \widetilde{T}
    \qty(2 g  T^2 - \frac{\hbar k_\text{L}}{2m} \widetilde{T} - 2 v_0  T) \label{eq: Analytical Phase Formula}
\end{align}
with $\widetilde{T} = \qty(2 + N^\text{loop}_\text{LMT}) T + N^\text{loop}_\text{LMT} T_\text{free}$.
Inserting the initial velocity $v_0$ from Eq.~\eqref{eq:init_velo} then yields the expression
\begin{align}
    \Delta \Phi_\text{Analytical} &= \gamma N_\text{loop} \frac{\hbar k_\text{L}^2}{m} T \widetilde{T}
    \left({N^\text{loop}_\text{LMT}}^2T_F-(2+N^\text{loop}_\text{LMT}) T\right).
    \label{eq: Analytical Phase Formula simplified}
\end{align}

\begin{figure}
    \centering
    \includegraphics[width=\linewidth]{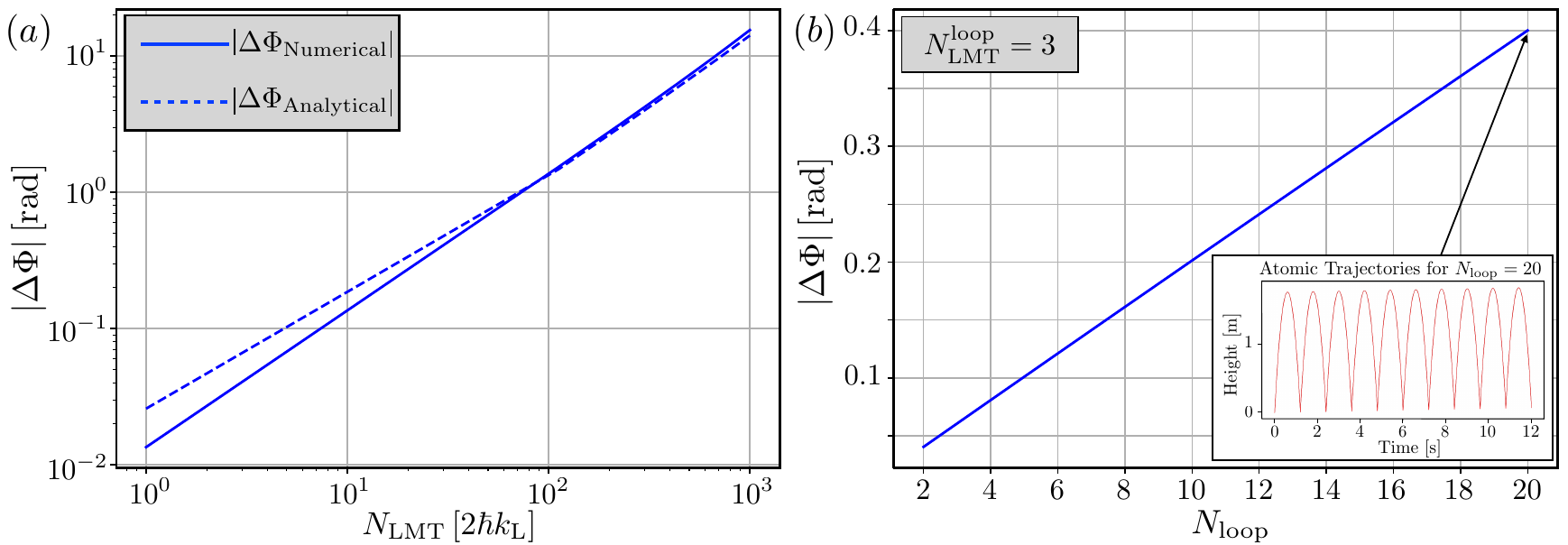}
    \caption{(a) Comparison of the analytical phase magnitude $\abs{\Delta \Phi_\text{Analytical}}$ from Eq.~\eqref{eq: Analytical Phase Formula} (dashed line) and a numerical Python simulation of the first order perturbed propagation phase magnitude $\abs{\Delta \Phi_\text{Numerical}}$ (solid line), as outlined in~\cite{phase_calculation}, as a function of $N^\text{loop}_\text{LMT}$ for two loops. (b) Numerical phase scaling with the number of loops for a fixed number of LMT pulses. Inset: Atomic trajectories for nine relaunches, i.e. 20 loops, each with $N^\text{loop}_\text{LMT} = 3$.}
    \label{fig: Phase_Numerics_vs_Analytics}
\end{figure}

For the case of sequential Bragg-type accelerations, we use the same initial velocity $v_0$ from Eq.~\eqref{eq:init_velo} and---for more than 2 loops---we relaunch both paths of the interferometer after every second loop with $v_\text{relaunch}^\text{SBD}$, which is given by
\begin{align}
    v_\text{relaunch}^\text{SBD} = 4 g T - \alpha N^\text{loop}_\text{LMT} \frac{\hbar k_L}{m}
\end{align}
with $\alpha \approx 0.7$. Note that the relaunch velocities are not the same for BOs and SBDs, since one of the defining constraints of the the relaunch velocity is that the atomic trajectories realign at zero after two loops. In the SBD case, as compared to the BO case, the atoms traverse a greater height, which leads to a smaller relaunch velocity.

A comparison between the analytical and numerical phase shifts obtained from the aforementioned approaches is shown in Fig.~\ref{fig: Phase_Numerics_vs_Analytics}\,(a), demonstrating good agreement between the analytical Bloch-type accelerations and the numerical sequential Bragg-type accelerations. 
The linear phase scaling with the number of loops and the atomic trajectories with periodic relaunches for a small number of LMT per loop can be seen in Fig.~\ref{fig: Phase_Numerics_vs_Analytics}\,(b). Note that the phase scales quite sensitively with the magnitude of the relaunch, especially when one considers thousands of relaunches. Higher than linear $N_\text{loop}$ scaling can be achieved, however, one needs to make sure that the atomic trajectories are always contained within the given baseline. 
The inset in Fig.~\ref{fig: Phase_Numerics_vs_Analytics}\,(b) additionally demonstrates the relaunching scheme for ten double-loops, which consist of nine relaunches.

\bibliographystyle{apsrev4-2}
\bibliography{references}